\documentstyle[12pt]{article}
\makeatletter
\newcommand{\singlespacing}{\let\CS=\@currsize\renewcommand{\baselinestretch}
{1.0}\tiny\CS}
\newcommand{\doublespacing}{\let\CS=\@currsize\renewcommand{\baselinestretch}
{1.5}\tiny\CS}

\begin{document}

\begin{center}
{\Large {\bf NONCOMPLEMENTARY WAVE--PARTICLE PHENOMENA REVISITED}}

\vspace{0.2cm}
{\bf G. Kar$^1$, A. Roy$^1$, S. Ghosh$^1$} and {\bf D. Sarkar$^2$}

$^1${\bf {Physics and Applied Math. Unit\\
          Indian Statistical Institute\\
          203, B. T. Road\\
          Calcutta--700 035\\
          India\\
          \underline{E--mail : res9708@isical.ac.in}}}

 $^2${\bf {Department of Mathematics\\
Barrackpore Rastraguru Surendranath College\\
85, Middle Road\\
Barrackpore, 24 Parganas (North)\\
West Bengal\\
India}}

\vspace{0.2cm}
{\bf ABSTRACT}
%\end{center}

\vspace{0.3cm}
The simultaneous verification of wave and particle property in some recently
suggested experiments has been reviewed in the light of Hilbert space
formalism. In this respect, the recent analysis of biprism experiment [J. L.
Cereceda, {\it Am. J. Phys.} {\bf 64} (1996) 459] is criticized. 

\end{center}

\newpage
\section{\bf Introduction}
For past few years there had been renewed discussion and critical comments on 
Bohr's complementarity. There have been suggested some new experiments which 
seems to challenge the wave particle duality ({\it i.e.} wave--particle
complementarity) as was suggested by Bohr [1]. Actually Bohr did not precisely 
formulate the principle of complementarity and its relation to quantum mechanics. 
This led to many controversies and misunderstanding, particularly in the case 
of complementarity of particle and wave properties. Recently many new experiments 
have been proposed and some of them have been performed where it was shown 
that same experimental arrangement can exhibit both particle and wave property.
In this respect the biprism experiment proposed by Ghose et. al. [2] and some 
other nice experimental arrangements proposed by Rangwala et. al. [3] are 
worth mentioning. Rangwala et. al. have clearly mentioned that the kind of 
wave and particle behaviour appearing in their arrangement are noncomplementary. 
D. Sen et. al. [4], broadly classified the well known complementary observables 
in two groups, where the example of the biprism experiment belongs to none.
Recently, Holladay [5] has given a restricted version of Bohr's wave--particle
complementarity, called ``which value--interference" complementarity
(which exists within the ``kinematic--dynamic complementarity," an example of
this being the complementarity of position (kinematic) and momentum (dynamic)
variables), and shown that the results of the above--mentioned biprism 
experiment violate the {\it usual} wave--particle complementarity of Bohr (in a rough
sense, which excludes the existence of wave and particle properties in a {\it
same} experimental arrangement), but do not violate the which
value--interference complementarity. But, simply by taking some restricted
version, we can not avoid the question of exact formulation ({\it i.e.}
mathematical formulation in Hilbert space) of complementarity
principle in quantum mechanics.  

A rigorous and excellent mathematical formulation of Bohr's complementarity
principle has been discussed by Busch and Lahti [6], in the context of Hilbert
space description of quantum mechanics. Here we shall briefly discuss this
formalism of Busch and Lahti in the context of wave--particle duality, and with
the help of this, we then analyse the aforesaid experiments. We will show that 
in all this experiments {\it the observables corresponding to particle and 
wave properties are commuting and hence deserve to be unambiguously verified in 
a single experimental arrangement}. In this light we will also discuss the 
analysis of biprism experiment by Cereceda [7], where he showed that the wave
and particle properties were indeed appearing simultaneously, and hence the
results of the biprism experiment violated Bohr's wave--particle complementarity. 
We argue that this analysis [7] is erroneous as it applies equation of 
complementarity (equation (15) of [7]) to a situation where there is no
complementarity (otherwise, rest of the paper [7] is a valuable exposition of
the wave--particle complementarity). More importantly, {\it equating} 
complementary principle to superposition principle [7], creates more confusion 
in understanding complementary principle. One should note that the three 
principles in Quantum Mechanics, namely superposition, uncertainty and
complementarity, {\it have only the common feature that each of them needs
noncommutative propositional structure for their description}, but otherwise
{\it none of them imply other} [8].  

In section 2, we shall give a brief account of the mathematical formalism of
Bohr's complementarity principle (in the Hilbert space description of quantum
mechanics), discussed by Busch and Lahti [6]. In section 3, we shall discuss
the proposed experiment on wave--particle properties by Rangwala and Roy [3] in
the light of the formalism described in section 2. In section 4, we shall
analyse of biprism experiment by Ghose, Home and Agarwal [2] in
the similar fashion of section 3, and compare these two experiments. In
section 5, we shall consider Cereceda's analysis of the biprism experiment [7].
And in section 6, we draw the conclusion.

\section{\bf Mathematical formalism}  
Here we introduce the notion of complementarity in the light of Hilbert space
formalism. In quantum mechanics any physical quantity is represented by self
adjoint operator $A$ on a seperable complex Hilbert space $H$, associated
to the system. The spectral measure of $A$ is denoted by $P_{A} : B(I\!\! R) \rightarrow P(H)$,
 where $B(I\!\! R)$ is the Borel $\sigma$--algebra of the real line $I\!\! R$  and $P(H)$
       is the set of all projection operators on $H$. Hence in quantum mechanics
measurement of any observable can be reduced to yes--no experiments of
propositions represented by projection operators. Any state in quantum mechanics
is represented by a positive bounded linear operator $T : H \rightarrow H$,
of trace one ({\it i.e.} ${\rm tr}T = 1$).   

Let $A$ and $B$ be two observables and $P_A(X)$, $P_B(Y)$ be their respective
projection operators (on $H$) with value sets $X$ and $Y$ respectively (thus
here $X$, $Y$ are Borel sets, {\it i.e.} they are  elements of $B(I\!\! R)$). 
Thus ${\rm tr}[TP_A(X)]$ is the probability that a measurement of the 
observable $A$ leads to a result in the value set $X$ when the system is 
prepared in the state $T$. 

Then \underline{the observables $A$ and $B$ are complementary if experimental}\\ 
\underline{arrangement for measuring $P_A(X)$ and that for $P_B(Y)$ are mutually}\\ 
\underline{exclusive for any bounded $X$ and $Y$, with none of $P_A(X)$, $P_B(Y)$}\\ 
\underline{being zero or identity operators}. The experimental arrangements for 
measuring $P_A(X)$ and that for $P_B(Y)$ are ``mutually exclusive" in the sense 
that a common measuring arrangement does not exist by which one can measure (with 
sharp values) simultaneously $P_A(X)$ and $P_B(Y)$.

The direct mathematical consequences of this result are :\\ 
\noindent{(i) $A$, $B$ are complementary if the greatest lower bound of $P_A(X)$ 
and $P_B(Y)$ is zero; or in other words, the closed subspaces corresponding to 
$P_A(X)$ and $P_B(Y)$ are disjoint.}\footnote{For any projection operator 
$E : H \rightarrow H$, the closed subspace (of $H$) corresponding to $E$ is 
the range $E(H)$ of $E$, where $H = E(H) + E(H)^{\perp}$, $E(H)^{\perp}$ being 
the closed subspace of $H$, orthogonal to $E(H)$.}
    
\noindent{(ii) In terms of probability, $A$, $B$ are complementary if for some 
(pure) state $\Psi (\in H)$ (thus here $T = |\Psi \rangle \langle \Psi|$), we 
have ${\rm tr}[TP_A(X)] \equiv \langle \Psi|P_A(X) \Psi \rangle = 1$, then 
we must have ${\rm tr}[TP_B(Y)] \equiv \langle \Psi| P_B(Y) \Psi \rangle < 1$ 
for all bounded value sets $X$, $Y$ [9].}\footnote{For {\it sharp} observables,
(i) and (ii) are equivalent; but for {\it unsharp} observables, (i) implies
(ii) [6]. Here we shall take (i) as the definition of complementary observables
$A$ and $B$.} 

Now considering the Hilbert space description of a quantum mechanical system,
to invalidate Bohr's complementarity principle, one has to be assured that
{\it every} pair of observables (associated with the system) are
noncomplementary, where two observables $A$ and $B$ are said to be
``noncomplementary" if at least one condition in the above--mentioned defition
(i) of complementarity is violated. It is to be mentioned that the {\it necessary} 
condition for $A$ and $B$ to be complementary :\\ 
they are totally noncommuting, {\it i.e.} having no eigenvector in
common.\\ 
So complementarity implies that the (closed) subspaces corresponding to 
$P_A(X)$ and $P_B(Y)$ are disjoint without being orthogonal to each other.
\underline{Two orthogonal projection operators} (hence their corresponding 
closed subspaces are also orthogonal, and so these subspaces also disjoint) 
\underline{always commute and hence noncomplementary.}

This will be our {\it main point} in analysing the various experiments showing 
simultaneously wave and particle properties.

\section{\bf The experiment of Rangwala and Roy}
Consider now the proposed experiment (figure (1)) as suggested by Rangwala and
Roy [3].

\begin{center}
\setlength{\unitlength}{1cm}
\begin{picture}(15,7)
\thicklines
\put(0,0){\framebox(3.0,1.5){\tiny{\rm{Single photon source}}}}
\put(3,0.75){\vector(1,0){1.5}}
\put(4.5,0.35){\makebox(0,0){$\Psi$}}
\put(4.5,0.75){\line(1,0){1.5}}
\put(6,0.75){\line(1,1){0.5}}
\put(6,0.75){\line(-1,-1){0.5}}
\put(6,0.75){\vector(1,0){1.5}}
\put(6,0.75){\vector(0,1){1.5}}
\put(5.65,0){\makebox(0,0){BS$_1$}}
\put(6,2.25){\line(0,1){3}}
\put(5.65,2.25){\makebox(0,0){${\Psi}_r$}}
\put(7.5,0.35){\makebox(0,0){${\Psi}_t$}}
\put(7.5,0.75){\line(1,0){1.5}}
%\put(6,3.75){\line(1,1){0.5}}
%\put(6,3.75){\line(-1,-1){0.5}}
%\put(5.65,3){\makebox(0,0){M$_1$}}
\put(5.75,5.25){\framebox(0.5,0.5){\tiny {D$_r$}}}
\put(9,0.75){\vector(1,0){1.5}}
\put(9,0.75){\line(1,1){0.5}}
\put(9,0.75){\line(-1,-1){0.5}}
\put(9,0.75){\vector(0,1){1.5}}
\put(8.65,0){\makebox(0,0){BS$_2$}}
\put(10.5,0.35){\makebox(0,0){${\Psi}_{tt}$}}
\put(10.5,0.75){\line(1,0){1.5}}
\put(8.65,2.25){\makebox(0,0){${\Psi}_{tr}$}}
\put(9,2.25){\line(0,1){1.5}}
\put(11.6,0){\makebox(0,0){M$_2$}}
\put(8.6,3){\makebox(0,0){M$_1$}}
\put(8.65,3.4){\line(-1,1){0.36}}
\put(8.825,3.575){\line(-1,1){0.36}}
\put(9,3.75){\line(-1,1){0.36}}
\put(9.175,3.925){\line(-1,1){0.36}}
\put(9.35,4.1){\line(-1,1){0.36}}
\put(12,0.75){\line(1,-1){0.36}}
\put(11.825,0.575){\line(1,-1){0.36}}
\put(11.65,0.4){\line(1,-1){0.36}}
\put(12.175,0.925){\line(1,-1){0.36}}
\put(12.35,1.1){\line(1,-1){0.36}}
\put(12,0.75){\vector(0,1){1.5}}
\put(12,2.25){\line(0,1){1.5}}
\put(12,0.75){\line(1,1){0.5}}
\put(12,0.75){\line(-1,-1){0.5}}
\put(9,3.75){\line(1,1){0.5}}
\put(9,3.75){\line(-1,-1){0.5}}
\put(9,3.75){\vector(1,0){1.5}}
\put(10.5,3.75){\line(1,0){1.5}}
\put(12,3.75){\line(1,1){0.5}}
\put(12,3.75){\line(-1,-1){0.5}}
\put(11.65,3){\makebox(0,0){BS$_3$}}
\put(12,3.75){\line(1,0){1.5}}
\put(12,3.75){\line(0,1){1.5}}
\put(13.5,3.5){\framebox(0.5,0.5){\tiny {D$_{t_2}$}}}
\put(11.75,5.25){\framebox(0.5,0.5){\tiny {D$_{t_1}$}}}

\end{picture}

\vspace{0.3cm}
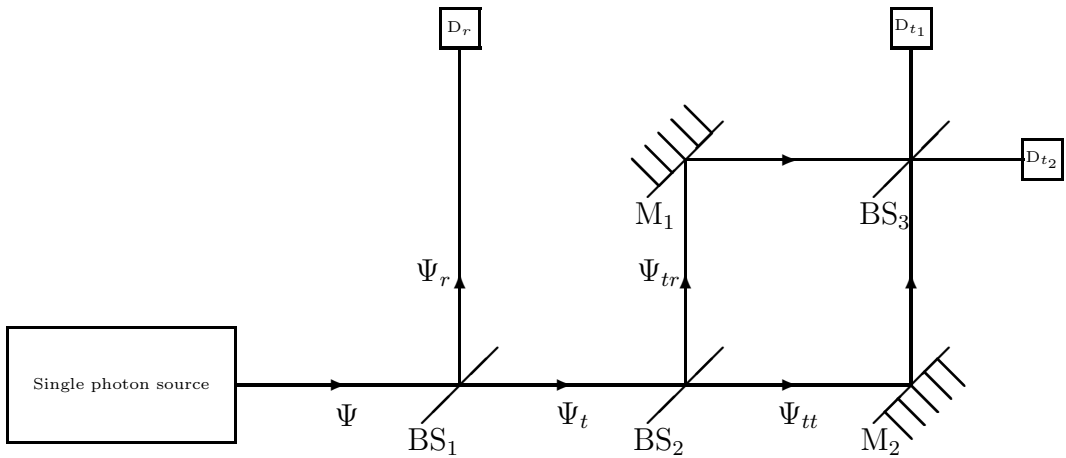
{\tiny Figure (1) : $\Psi$ is the initial single photon state; BS$_1$, BS$_2$, BS$_3$
are 50 : 50 beam splitters; M$_1$, M$_2$ are perfect reflecting mirrors;
${\Psi}_t$, ${\Psi}_r$ being respectively the transmitted and reflected paths
by BS$_1$; ${\Psi}_{tt}$, ${\Psi}_{tr}$ being respectively the transmitted and reflected paths
by BS$_2$; D$_r$, D$_{t_1}$ and D$_{t_2}$ are photon detectors.}
\end{center}

%\vspace{3cm} 

Here for the arrangement (fig. (1)), the associated Hilbert space is three 
dimensional with orthogonal basis $\left \{ {\Psi}_r, {\Psi}_{tr}, {\Psi}_{tt} \right \}$.

Now for a single photon incident on the beam--splitter (BS$_1$), there will be
anticoincidence between D$_r$ and D$_{t_1}$ (or D$_{t_2}$) and the distribution
of counts at D$_{t_1}$ and D$_{t_2}$ will show an interference pattern depending on the phase difference. We now construct the path and
interference observables.

The path observable corresponding to the detector D$_r$ is the projection
operator P{$\left [{\Psi}_r \right]$} and the path observable corresponding to
detection in either D$_{t_1}$ or D$_{t_2}$ is ${\rm P}{\left [{\Psi}_{tr}
\right]} + {\rm P}{\left [{\Psi}_{tt} \right]} = {\rm P}_{t_1 t_2}$ (say). The interference observables
for 50 : 50 beam--splitter BS$_3$, are represented by projection operators
${\rm }P{\left [\frac{1}{\sqrt 2} \left ({\Psi}_{tt} + {\Psi}_{tr} \right)
\right]}$ and ${\rm }P{\left [\frac{1}{\sqrt 2} \left ({\Psi}_{tt} - {\Psi}_{tr} \right)
\right]}$. Here P[.] is the projection operator on the vector inside the square
bracket.  

It is to be noted that both the interference observables are defined on the
subspace ${\rm S}_{t_1 t_2}$ spanned by ${\Psi}_{tr}$ and ${\Psi}_{tt}$, and ${\rm S}_{t_1 t_2}$ is
orthogonal to ${\Psi}_{r}$. Hence both the interference observables 
commute with ${\rm P}{\left [{\Psi}_{r} \right]}$. Commutativity of each of
these interference observables with the other path observable 
${\rm P}_{t_1 t_2}$ follows from the simple fact that both the vectors 
$\frac{1}{\sqrt 2} \left ({\Psi}_{tt} \pm {\Psi}_{tr} \right)$ are elements of 
${\rm S}_{t_1 t_2}$. Hence it follows that in this setup the concerned path 
and interference observables are {\it noncomplementary}. All the measurements 
discussed in [3], including the biprism experiment [2], belong to this class,
{\it i.e.}, {\bf the observable representing wave has its support contained
within the support of one of the path observables, and hence commutes with
both of these path observables}. 

\section{\bf The biprism experiment}
For clear understanding, let us discuss the biprism experiment (given in fig.
(2)) in the mathematical framework discussed in section 2. 
\begin{center}
\setlength{\unitlength}{1cm}
\begin{picture}(11,2)
\thicklines
\put(0,0){\framebox(3.0,1.5){\tiny{\rm{Single photon source}}}}
\put(3,0.75){\vector(1,0){1.5}}
\put(4.5,0.35){\makebox(0,0){$\Psi$}}
\put(4.5,0.75){\line(1,0){1.0}}
\put(6,0.75){\line(-1,1){0.5}}
\put(6,0.75){\line(1,-1){0.5}}
\put(5.5,0.75){\line(0,1){0.5}}
\put(5.5,0.75){\line(0,-1){0.5}}
\put(5.5,0.25){\line(1,0){1.0}}
\put(6.25,0.75){\line(-1,1){0.5}}
\put(6.25,0.75){\line(1,-1){0.5}}
\put(6.75,0.75){\line(0,1){0.5}}
\put(6.75,0.75){\line(0,-1){0.5}}
\put(6.75,1.25){\line(-1,0){1.0}}
\put(6.75,0.75){\vector(1,0){1.5}}
\put(6,0.25){\vector(0,-1){1.5}}
%\put(5.65,0){\makebox(0,0){BS$_1$}}
%\put(6,2.25){\line(0,1){3}}
%\put(5.65,2.25){\makebox(0,0){${\Psi}_r$}}
%\put(7.5,0.35){\makebox(0,0){${\Psi}_t$}}
%\put(7.5,0.75){\line(1,0){1.5}}
%\put(6,3.75){\line(1,1){0.5}}
%\put(6,3.75){\line(-1,-1){0.5}}
%\put(5.65,3){\makebox(0,0){M$_1$}}
\put(5.75,-1.75){\framebox(0.5,0.5){\tiny {D$_r$}}}
%\put(9,0.75){\vector(1,0){1.5}}
%\put(9,0.75){\line(1,1){0.5}}
%\put(9,0.75){\line(-1,-1){0.5}}
%\put(9,0.75){\vector(0,1){1.5}}
%\put(8.65,0){\makebox(0,0){BS$_2$}}
%\put(10.5,0.35){\makebox(0,0){${\Psi}_{tt}$}}
%\put(10.5,0.75){\line(1,0){1.5}}
%\put(8.65,2.25){\makebox(0,0){${\Psi}_{tr}$}}
%\put(9,2.25){\line(0,1){1.5}}
%\put(11.6,0){\makebox(0,0){M$_2$}}
%\put(8.6,3){\makebox(0,0){M$_1$}}
%\put(8.65,3.4){\line(-1,1){0.36}}
%\put(8.825,3.575){\line(-1,1){0.36}}
%\put(9,3.75){\line(-1,1){0.36}}
%\put(9.175,3.925){\line(-1,1){0.36}}
%\put(9.35,4.1){\line(-1,1){0.36}}
%\put(12,0.75){\line(1,-1){0.36}}
%\put(11.825,0.575){\line(1,-1){0.36}}
%\put(11.65,0.4){\line(1,-1){0.36}}
%\put(12.175,0.925){\line(1,-1){0.36}}
%\put(12.35,1.1){\line(1,-1){0.36}}
%\put(12,0.75){\vector(0,1){1.5}}
%\put(12,2.25){\line(0,1){1.5}}
%\put(12,0.75){\line(1,1){0.5}}
%\put(12,0.75){\line(-1,-1){0.5}}
%\put(9,3.75){\line(1,1){0.5}}
%\put(9,3.75){\line(-1,-1){0.5}}
%\put(9,3.75){\vector(1,0){1.5}}
%\put(10.5,3.75){\line(1,0){1.5}}
%\put(12,3.75){\line(1,1){0.5}}
%\put(12,3.75){\line(-1,-1){0.5}}
%\put(11.65,3){\makebox(0,0){BS$_3$}}
%\put(12,3.75){\line(1,0){1.5}}
%\put(12,3.75){\line(0,1){1.5}}
\put(8.25,0.5){\framebox(0.5,0.5){\tiny {D$_t$}}}
%\put(11.75,5.25){\framebox(0.5,0.5){\tiny {D$_{t_1}$}}}
\thinlines
\put(5.5,0.75){\line(1,0){0.5}}
\put(6,0.75){\line(1,0){0.75}}
\put(6,0.75){\line(0,-1){0.5}}
\end{picture}

\vspace{1.5cm}
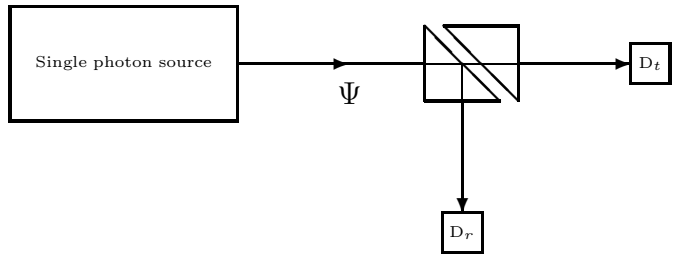
{\tiny Figure (2) : $\Psi$ is the initial single photon state; 
%BS$_1$, BS$_2$, BS$_3$ are 50 : 50 beam splitters; M$_1$, M$_2$ are perfect reflecting mirrors;
%${\Psi}_t$, ${\Psi}_r$ being respectively the transmitted and reflected paths
%by BS$_1$; ${\Psi}_{tt}$, ${\Psi}_{tr}$ being respectively the transmitted and reflected paths
%by BS$_2$; 
D$_r$, D$_t$ are photon detectors corresponding to reflected and transmitted 
photons.}
%and D$_{t_2}$ are photon detectors.}
\end{center}

It is to be mentioned that the minimum Hilbert space required for the proper 
description of a system must be clearly specified. For example, in the case of 
double--slit experiment two dimensional Hilbert space is sufficient wheras, 
the experiment described in fig. (1) requires three dimensional Hilbert space. 
In the case of the biprism experiment, the tunneling of the photon through the 
gap between the prisms is a manifestation of (and is {\it defined} as) some wave 
phenomenon, because tunneling depends on the relation between the gap and wave 
length of the photon. 
%So it is analogous to barrier potential problem in standard quantum mechanics, 
%as there also tunneling of quantum particle depends on the height and width of 
%the barrier.

Now for the description of this kind of wave phenomenon we need higher 
dimensional Hilbert space $H$, as shown in our previous example (figure (1)).
The wave phenomenon ({\it i.e.} tunneling) described in this biprism experiment
and the wave phenomenon ({\it i.e.} interference) described in the experiment
of fig. (1) have some similarity in nature [3] (and we shall see that, in the case of
the biprism experiment, simultaneous verification of wave and 
particle phenomena follows from commutativity of the corresponding observables
-- the same thing also happens in the experiment of fig. (1)). So for the 
proper description of the biprism experiment, we need a (higher dimensional) 
Hilbert space $H$ where the closed subspace (of $H$) generated by all possible 
reflected paths is $H_r$ (of dimension more than one) and similarly $H_t$ (of
dimension more than one) is defined for transmitted wave.\footnote{Reflection in fig. (2) is also a manifestation of wave property,
because reflection (actually it is internal reflection) here depends on the
relation between the prism--gap and the wave length of the photon.} In general, 
$H_r$ and $H_t$ are infinite dimensional. Here, $H = H_r \oplus H_t$.

The suggested wave property (tunneling) in biprism experiment, must be represented 
by some projection operator ${\rm P}_{wave}$ (say) (defined on $H$), where
${\rm P}_{wave}(H)$ is contained in the subspace $H_t$. So 
${\rm P}_{wave} \le {\rm P}_t$ ({\it i.e.} $\langle \psi| {\rm P}_{wave} |\psi
\rangle \le \langle \psi| {\rm P}_t |\psi \rangle$ for every $|\psi \rangle$ in
$H$), where ${\rm P}_t$ is projection operator on $H_t$ ({\it i.e.} ${\rm
P}_t(H) = H_t$).\footnote{Now in fig. (2), it is clear by the definition of the
wave property (which is tunneling here) that ``tunneling of photon through the
prism--gap" implies ``detection of that photon at the detector ${\rm D}_t$,"
and ``detection of a photon at the detector ${\rm D}_t$" implies ``tunneling of
that photon". So here ${\rm P}_{wave} = {\rm P}_t$, while in general (as in the
case of fig. (1)), ${\rm P}_{wave} \le {\rm P}_t$. And in this respect, the
experiments described in figures (1) and (2) differ.} Then obviously 
${\rm P}_{wave}$ commutes with both the path observables ${\rm P}_r$ (where
${\rm P}_r$ is the projection operator on $H$ with ${\rm P}_r(H) = H_r$) 
%\left[{\Psi}_r \right]$ 
and ${\rm P}_t$, and their simultaneous verification is not unwarrented. There 
is no complementarity, {\it i.e.}, verification of wave property in this example 
{\it implies} verification of the transmmited path property.

\section{\bf The analysis of Cereceda} 
Cereceda, in his analysis of complementarity [7], has taken the relation 
\begin{equation}
{\rm P}^2 + {\rm W}^2 = 1
\end{equation}
as the equation of complementarity where ${\rm P}$ is some measure of path
information, and ${\rm W}$ is the visibility of interference fringes; or in
general, ${\rm P}$ is a measure of particle property and ${\rm W}$ corresponds
to wave property. We think that the relation (1) should be judged in its proper
perspective. In the case of unsharp joint measurement of path and interference,
as suggested by Wooters and Zurek [10] and many others ([11], [12]), and even in
the case described by Cereceda in figure (3) of [7], this relation (i.e. equation
(1) ) expresses complementarity phenomena in some form; but extension of this
relation directly to biprism experiment (where there is no complementarity) and
more importantly, to equate it ({\it i.e.} equation (1)) to superposition 
principle, is erroneous.

In Cereceda's analysis [7], the state of the single-photon emerging from the biprism
arrangement is expressed in the form
\setcounter{equation}{1}
\begin{equation}
\Psi\, = \,\alpha\Psi_r + \beta\Psi_t\, ,
\end{equation}
where the coefficients $\alpha$ and $\beta$ fulfill the relation $|\alpha|^2+|\beta|^2
=1$, provided the prisms are lossless. In the superposition state (2), $\Psi_r$
corresponds to reflected wave and $\Psi_t$ corresponds to transmitted wave; $\Psi_r$
and $\Psi_t$ are orthogonal. This means that the two detectors $D_r$ and $D_t$ in
Fig. 2 should click in perfect anticoincidence for true single-photon incident states,
thus showing unambiguous particle-like propagation of the detected photons [13].
Actual transmitted path property in biprism experiment corresponds to a projection
operator ${\rm P}_t\::\: H \rightarrow H_t$ (where, as explained in Section 4, $H_t$
is in general an infinite dimensional closed subspace of $H$ (the Hilbert space of the
system described), with ${\rm P}_t(H)=H_t$). Analogously, reflected path property
corresponds to a projection operator ${\rm P}_r\::\: H \rightarrow H_r$, with $H_r = H_t^{\perp}$. The important point to be emphasized here is that the defined wave
property ({\it{i.e.}} tunneling), to which the projection operator ${\rm P}_{wave}$ corresponds, is completely defined in the subspace $H_t$ ({\it{i.e.}} ${\rm P}_{wave}
(H)$ is contained in $H_t$). In physical terms, this simply means that a photon
tunneling through the gap (thereby showing a wave-like behavior) does also entail
which-path information of that photon toward detector $D_t$ (that is, transmitted path property).

In the analysis of the biprism experiment expounded in Ref. 7, the quantity $|\alpha|$
is regarded as the amount of which-path information, whereas the quantity $|\beta|$ is
regarded as the amount of wave information. This interpretation, however, turns out to
be oversimplified since, as we have said, the transmission (tunneling) amplitude
$\beta$ provides which-path information as well, as soon as the photon is detected by
$D_t$. As a result, the purported equivalence of the complementarity relation (1) with
the normalization of the superposition state (2) cannot be maintained. Let us examine
this issue in the light of the Hilbert space description of quantum-mechanical systems.
In its true representation, $\Psi_t \in H_t$ and $\Psi_r \in H_r(=H_t^{\perp})$, so
that the following relations $|\Psi_r\rangle \langle\Psi_r| \leq {\rm P}_r$ and
$|\Psi_t\rangle \langle\Psi_t| \leq {\rm P}_t$ generally apply. Consider now the
simplest (idealized) situation where $|\Psi_r\rangle \langle\Psi_r| = {\rm P}_r$ and

$|\Psi_t\rangle \langle\Psi_t| = {\rm P}_t$. So, if $|\alpha|$ can be regarded as
measure of (reflected) path property, we must have $|\alpha|=\sqrt{\langle\Psi|
{\rm P}_r|\Psi \rangle}$. Similarly, if $|\beta|$ is a (defined) wave property then
it can be expressed in the form $|\beta|=\sqrt{\langle\Psi| {\rm P}_{wave}|\Psi
\rangle}$. In addition to this, however, $|\beta|$ has to be regarded as alternative (transmission) path property (as $\sqrt{\langle\Psi| {\rm P}_{wave}|\Psi \rangle} \leq \sqrt{\langle\Psi| {\rm P}_t|\Psi \rangle}$\,), and thus the normalization condition $|\alpha|^2+|\beta|^2=1$ here in no way manifests any complementary phenomenon.\footnote{Formally, since ${\rm P}_{wave}={\rm P}_t$ in the biprism experiment, and in the
idealized situation we are considering, we may replace ${\rm P}_{wave}$ with
${\rm P}_t$ to obtain $|\alpha|^2+|\beta|^2= \langle\Psi|{\rm P}_r|\Psi \rangle + \langle\Psi|{\rm P}_{wave}|\Psi \rangle = \langle\Psi|{\rm P}_r|\Psi \rangle + 
\langle\Psi|{\rm P}_t|\Psi \rangle = 1$. Nevertheless, any resemblance of this
expression to the complementarity relation ${\rm P}^2+{\rm W}^2=1$ is merely
coincidental, as the transmission amplitude $\beta$ in the biprism experiment
necessarily entails both wave and particle property.} In fact, as was stated in
Section 4, there is no complementarity, and then both classical wave and particle
pictures must be invoked at the same time in order to account for the simultaneous
verification of the mutually {\em noncomplementary} tunneling and transmitted path
properties (or else, internal refection and reflected path properties).

\section{\bf Conclusion} 
In conclusion, the noncomplementary nature of wave and particle property in the biprism experiment (and some other
experiments of this category), follows from the commutativity, when described
in the Hilbert space formalism of quantum mechanics. And hence, the results of
these experiments, in no way, violate Bohr's complementarity principle, the
form we have taken here.       

{\noindent {\bf Acknowledgement :}} The authors are thankful to Prof. J. L.
Cereceda for passing his valuable comments in rewritting the paper,
particularly section 5.

\newpage

\vspace{0.4cm}
\begin{center}
{\bf REFERENCES}
\end{center}

\vspace{0.3cm}
\begin{enumerate}
\item N. Bohr, {\it Naturwissenschaften} {\bf 16} (1928) 245.
\item P. Ghose, D. Home and G. S. Agarwal, {\it Phys. Lett. A} {\bf 153} (1991)
403; {\it ``Conceptual Foundation of Quantum Mechanics"} by D. Home (1997,
Plenum Press, N. Y.) ch. 5.
\item S. Rangwala and S. M. Roy, {\it Phys. Lett. A} {\bf 190} (1994) 1.
\item D. Sen, A. N. Basu and S. Sengupta, {\it Helv. Phys. Acta} {\bf 67}
(1994) 755.
\item W. G. Holladay, {\it Am. J. Phys.} {\bf 66} No. 1 (1998) 27.
\item P. Busch and P. J. Lahti, {\it Riv. Del Nuovo Cimento} {\bf 18} No. 4,
(1995) 1.
\item J. L. Cereceda, {\it Am. J. Phys.} {\bf 64} (1996) 459.
\item P. J. Lahti, {\it Int. J. Theor. Phys.} {\bf 29} (1980) 905.
\item {\it ``The Logic of Quantum Mechanics"} by E. G. Beltrametti and G.
Cassinelli (1981, Addison--Wesley).
\item W. K. Wooters and W. Zurek, {\it Phys. Rev. D} {\bf 19} (1979) 473.
\item D. M. Greenberger and A. Yasin, {\it Phys. Lett. A} {\bf 128} (1988) 391.
\item B. -G. Englert, {\it Phys. Rev. Lett.} {\bf 77} (1996) 2154.
\item P. Grangier, G. Roger, and A. Aspect, {\it Europhys. Lett.} {\bf 1} (1986) 173.
\end{enumerate}

\end{document}